\documentclass[pdflatex,sn-mathphys-num]{sn-jnl}
\usepackage[T1]{fontenc}
\usepackage[utf8]{inputenc}
\usepackage{amsmath,amssymb,amsfonts}
\usepackage{graphicx}
\usepackage{booktabs,longtable,array,calc}
\usepackage{xurl}
\providecommand{\tightlist}{\setlength{\itemsep}{0pt}\setlength{\parskip}{0pt}}
\providecommand{\pandocbounded}[1]{#1}
\hypersetup{hypertexnames=false,pdftitle={Query Language Sensitivity in Automated Sentiment Scores of LLM
Responses},pdfauthor={Dmitrij Żatuchin}}
\begin{document}
\title[Query Language Sensitivity in Automated Sentiment Scores]{Query Language Sensitivity in Automated Sentiment Scores of LLM
Responses}
\author*[1,2]{\fnm{Dmitrij} \sur{\.Zatuchin}}\email{dmitrij.zatuchin@eek.ee}
\affil*[1]{\orgdiv{Department of Information Technologies}, \orgname{Estonian Entrepreneurship University of Applied Sciences (EUAS)}, \orgaddress{\city{Tallinn}, \country{Estonia}}}
\affil[2]{\orgname{Rankfor.AI Sp. z o.o.}, \orgaddress{\city{Wroc\l{}aw}, \country{Poland}}}
\abstract{Automated scores are often used to summarize how large language models
describe brands across prompts, providers, and languages. Such scores
can vary because the answers differ, because the scoring instrument
behaves differently across texts or languages, or for both reasons. We
examine this measurement problem in 12,933 responses from a design
crossing 20 stored brand labels, eight query languages, three provider
aliases, and 15 prompt identifiers. The outcome is the signed
winning-class confidence from a multilingual sentiment classifier:
neutral winners are coded zero, while positive and negative winners
receive signed confidence. A crossed linear mixed model fitted to all
responses assigns 26.49\% of fitted variance to query language and
1.46\% to brand. An exploratory refit after excluding English assigns
5.07\% to language and 2.23\% to brand. These percentages have different
denominators: total fitted variance falls by 65.08\%, and absolute brand
variance falls by 46.65\%. English contains 591 negative classifier
scores among 1,620 answers, compared with 77 among 11,313 answers in the
other seven languages. The result shows that the original variance
decomposition is highly sensitive to English inclusion. It does not
identify a causal English effect or establish that English answers are
more negative toward the brands. The classifier was trained for tweet
sentiment, receives no separate target-brand condition, and English is
the only study language represented in its documented sentiment
fine-tuning set. Automated score differences should therefore be checked
as properties of both the answers and the measurement instrument before
they guide substantive conclusions.}
\maketitle
\noindent\emph{Revised analysis of \href{https://arxiv.org/abs/2607.13304v1}{arXiv:2607.13304v1},
prepared 1 November 2026.}

\section{Introduction}\label{introduction}

Studies of brand representation in large language model (LLM) answers
need a response-level outcome. Researchers may code whether a brand is
named or recommended, record its position in a list, compare answer
embeddings, or apply a text classifier. Each choice defines a different
quantity. A sentiment classifier is attractive because it produces one
numeric score for every answer, including answers in several languages.
Its availability does not guarantee that the resulting scores are
comparable across those languages or that the score refers specifically
to the target brand.

This paper revisits a variance-components analysis of an automated
sentiment score. The original analysis asked how much variation in one
scored LLM answer was associated with brand, query language, provider,
prompt, and the remaining residual. The point estimates reproduce. A
subsequent sensitivity analysis shows that the main language component
depends strongly on the inclusion of English. English accounts for 591
of the 668 negative classifier scores, even though it accounts for 1,620
of 12,933 answers. When English is removed, the fitted language share
falls from 26.49\% to 5.07\%.

That change requires careful interpretation. The two percentages come
from different fitted populations and different total-variance
denominators. Removing English changes the observations and the language
population, and every variance component is re-estimated. The comparison
therefore shows sensitivity to sample composition. It does not partition
the original language component into an English contribution and a
remainder. It also cannot distinguish genuine differences in answer
content from unequal behaviour of the scoring instrument.

The measurement concern is substantive. The scorer is
\nolinkurl{cardiffnlp/twitter-xlm-roberta-base-sentiment-multilingual}, a
multilingual transformer fine-tuned for tweet sentiment {[}1-3{]}. Its
documented sentiment fine-tuning set covers eight languages, of which
English is the only language that overlaps with this study's eight query
languages {[}4{]}. The available scoring implementation uses the first
2,000 answer characters, applies token truncation at 512, assigns
exactly zero when the neutral class wins, and otherwise assigns the
winning positive or negative confidence with its corresponding sign. It
receives the answer text without a separate target-brand input. These
properties make the outcome a signed classifier score for the response,
not a validated measure of sentiment toward the target brand.

The corrected contribution is focused. First, we state the automated
outcome and its provenance precisely. Second, we reproduce the original
main-effects variance decomposition. Third, we report an exploratory
English-exclusion refit and the accompanying change in both shares and
absolute variances. Fourth, we set out the conclusions that the data
support and the validation needed for claims about actual brand
sentiment. We do not use this analysis to rank brands, make a binary
recommendation claim, or prescribe how a query budget should be
allocated.

\section{Measurement background}\label{measurement-background}

\subsection{Variation in LLM outputs and variation in their
scores}\label{variation-in-llm-outputs-and-variation-in-their-scores}

Repeated LLM requests can yield different outputs even when generation
settings appear fixed {[}5,6{]}. In a multilingual brand study,
additional variation can arise from the provider system, prompt task,
query language, and target brand. A crossed design can estimate how the
variance of a numeric outcome is distributed across these facets.
Generalizability theory and linear mixed models provide established
frameworks for such decompositions {[}7-11{]}.

Two neighbouring studies use different outcomes. The synthesised
consideration set measures brand inclusion and concentration across
recommendation categories {[}12{]}, with corrected analyses and
replication outputs archived separately {[}14{]}. The Dice Roll Method
specifies a repeated-query audit protocol for brand recommendations and,
in version 2, withdraws fixed reliability tiers {[}13{]}. Neither study
validates the sentiment score analysed here. Their inclusion and
recommendation outcomes should remain separate from a response-level
classifier score.

A variance decomposition remains conditional on its outcome. Here, the
outcome is an automated response-level sentiment score. A language
component in that score can reflect at least two processes. The
underlying answers may differ by language. The scorer may also map
linguistically or structurally similar content to different values
across languages. Domain shift can add another source of measurement
difference because the classifier was fine-tuned on tweets while the
study answers are often long and list-structured. Truncation can change
which part of an answer is scored. These mechanisms can operate
together.

The variance component for language is therefore a property of the
scored responses under this collection and scoring pipeline. It cannot,
by itself, be interpreted as a general language effect on brand
reputation. The same restriction applies to the brand component: it
measures variation associated with the study's brand labels in this
outcome, not the accuracy of a brand ranking or recommendation measure.

\subsection{The scorer and language
overlap}\label{the-scorer-and-language-overlap}

The scorer combines the XLM-R multilingual encoder with a Twitter
sentiment fine-tune {[}1-3{]}. The checkpoint model card points to the
multilingual tweet sentiment dataset, whose documented languages are
Arabic, English, French, German, Hindi, Italian, Portuguese, and Spanish
{[}4{]}. The study languages are Czech, English, Estonian, Hungarian,
Lithuanian, Latvian, Polish, and Slovak. English is the sole overlap.

This mismatch does not prove that the classifier fails in the other
seven languages. XLM-R pretraining covers many languages, and
cross-lingual transfer is one of its intended capabilities. It does mean
that equal calibration for this study's languages and domain cannot be
assumed from the fine-tuning documentation. A substantive comparison of
company sentiment requires validation against labels created for the
target construct, across the languages and answer forms being compared.

\section{Data and design}\label{data-and-design}

\subsection{Corpus}\label{corpus}

The analysis uses 12,933 retained response records with labels for 20
selected brands operating in Central and Eastern European markets. The
source archive records collection from 31 March to 10 April 2026. The
response-level corpus underlies the aggregate cross-language release
{[}15{]}; the public aggregate release does not supply these individual
rows. The brands span several industries and recognition levels; the
sampling description does not imply that every company is headquartered
in the region. Questions were issued in eight languages and routed
through three provider aliases: \texttt{openai}, \texttt{gemini}, and
\texttt{perplexity}. The response records preserve provider aliases, not
immutable runtime model versions. Collection materials report a
temperature setting of 0.3, but complete per-request technical receipts
are unavailable. We therefore do not assign unverified runtime model
versions to individual responses.

The design contains 15 prompt identifiers in five categories. Twelve
identifiers were collected once per intended brand, language, and
provider cell, yielding 5,760 single responses. The three Category D
identifiers were repeated within cells. D1 requests a description, D2
asks for strengths and weaknesses, and D3 asks whether the system
recommends the named brand. They are different tasks, not
interchangeable paraphrases. Category D contains 7,173 retained
responses in 1,435 observed cells: 1,433 cells have five responses and
two have four. Five scheduled cells have no usable response. The
retained total is therefore 5,760 plus 7,173, or 12,933.

All fifteen prompt identifiers enter the main model. Some categories
explicitly name the target brand. Category E contains unbranded industry
questions; its stored target-brand label organizes the original
collection and does not establish that the brand appeared in the prompt
or answer. The present outcome does not encode target-name mention,
endorsement, rank, or recommendation.

\subsection{Outcome construction}\label{outcome-construction}

The available implementation scores the first 2,000 characters of each
response with
\nolinkurl{cardiffnlp/twitter-xlm-roberta-base-sentiment-multilingual}; the
tokenizer truncates to 512 tokens. Let the classifier's winning class be
negative, neutral, or positive, with winning-class confidence \(c\). The
available implementation defines the score as

\[
s = \begin{cases}
-c, & \text{if the negative class wins},\\
0, & \text{if the neutral class wins},\\
c, & \text{if the positive class wins}.
\end{cases}
\]

A complete per-response scoring execution trace is unavailable. This
correction analyses the archived numerical scores and does not rerun the
classifier. The available scorer receives no separate target-brand
condition. A negative value means that the negative response-level class
won. It does not establish hostility, a negative recommendation, or a
negative statement about the target brand. Of the 12,933 scores, 11,879
are exactly zero, or 91.85\%. The highly concentrated distribution makes
a homogeneous Gaussian residual model an approximation.

Table 1 reports the complete score-class distribution by stored
language. The source uses \texttt{cz} as the Czech label.

\textbf{Table 1. Stored classifier winning-class distribution by
language.}

{\def\LTcaptype{table} 
\begin{longtable}[]{@{}
  >{\raggedright\arraybackslash}p{(\linewidth - 8\tabcolsep) * \real{0.1579}}
  >{\raggedleft\arraybackslash}p{(\linewidth - 8\tabcolsep) * \real{0.2105}}
  >{\raggedleft\arraybackslash}p{(\linewidth - 8\tabcolsep) * \real{0.2105}}
  >{\raggedleft\arraybackslash}p{(\linewidth - 8\tabcolsep) * \real{0.2105}}
  >{\raggedleft\arraybackslash}p{(\linewidth - 8\tabcolsep) * \real{0.2105}}@{}}
\toprule\noalign{}
\begin{minipage}[b]{\linewidth}\raggedright
Stored language
\end{minipage} & \begin{minipage}[b]{\linewidth}\raggedleft
Responses
\end{minipage} & \begin{minipage}[b]{\linewidth}\raggedleft
Negative
\end{minipage} & \begin{minipage}[b]{\linewidth}\raggedleft
Neutral (score 0)
\end{minipage} & \begin{minipage}[b]{\linewidth}\raggedleft
Positive
\end{minipage} \\
\midrule\noalign{}
\endhead
\bottomrule\noalign{}
\endlastfoot
cz & 1,610 & 3 & 1,603 & 4 \\
en & 1,620 & 591 & 1,018 & 11 \\
et & 1,620 & 15 & 1,391 & 214 \\
hu & 1,619 & 8 & 1,577 & 34 \\
lt & 1,620 & 1 & 1,562 & 57 \\
lv & 1,619 & 0 & 1,575 & 44 \\
pl & 1,620 & 27 & 1,576 & 17 \\
sk & 1,605 & 23 & 1,577 & 5 \\
\textbf{Total} & \textbf{12,933} & \textbf{668} & \textbf{11,879} &
\textbf{386} \\
\end{longtable}
}

English contributes 591 of the 668 negative scores, or 88.47\%, while
representing 12.53\% of responses. Within English, 591 of 1,620 answers
receive a negative score, or 36.48\%. In the other seven languages, 77
of 11,313 answers receive a negative score, or 0.68\%. These are
classifier outcomes, not human-coded evaluations of the answers.

\section{Statistical analysis}\label{statistical-analysis}

\subsection{Main-effects model}\label{main-effects-model}

Let \(y_{b\ell mp i}\) denote the stored classifier score for brand
\(b\), language \(\ell\), provider alias \(m\), prompt identifier \(p\),
and iteration \(i\). The main model is an intercept-only crossed linear
mixed model:

\[
y_{b\ell mp i} = \mu + u_b + u_\ell + u_m + u_p + \varepsilon_{b\ell mp i},
\]

where the random intercepts are mutually independent and Gaussian,

\[
u_b \sim N(0,\sigma_b^2),\quad
u_\ell \sim N(0,\sigma_\ell^2),\quad
u_m \sim N(0,\sigma_m^2),\quad
u_p \sim N(0,\sigma_p^2),\quad
\varepsilon \sim N(0,\sigma_e^2).
\]

The fitted total variance is

\[
V = \sigma_b^2 + \sigma_\ell^2 + \sigma_m^2 + \sigma_p^2 + \sigma_e^2,
\]

and the reported share for component \(j\) is \(\sigma_j^2/V\). Because
M1 contains main effects only, the residual includes within-cell
variation, interactions, and other unmodelled structure. It should not
be described as pure repeat noise.

The model was fitted by restricted maximum likelihood with one constant
grouping variable and crossed variance-component terms for brand,
language, provider, and prompt. The original implementation used
\texttt{statsmodels} and BFGS with a maximum of 3,000 iterations. It
contained no additional group random intercept. The preserved
full-corpus result and the independent reproduction have identical
loaded floating-point component values.

\subsection{Exploratory English-exclusion
sensitivity}\label{exploratory-english-exclusion-sensitivity}

The sensitivity analysis removes rows whose stored language is
\texttt{en} and refits the same M1 specification without recoding any
score. It is exploratory because the original analysis did not specify
English exclusion as a confirmatory test. The refit changes both the
observations and the language population. Comparisons between fitted
shares must therefore be accompanied by the absolute components and each
fit's total variance.

Both fits report convergence. The independent run retained warnings that
the original script suppressed. Each fit produced a
boundary-of-parameter-space warning and repeated numerical warnings
during optimization. These warnings do not negate the exact
point-estimate reproduction, but they limit confidence in uncertainty
calibration. No bootstrap intervals or alternative residual-distribution
model is reported here.

An additional exploratory check omits each of the other seven languages
in turn. Following a failed BFGS fit for Czech exclusion, one Powell
retry was introduced and retained alongside the failed attempt in the
diagnostics. The interpreted comparison uses the converged retry. These
refits are descriptive point estimates, with no confidence intervals or
significance tests.

\section{Results}\label{results}

Table 2 gives the point estimates for the original corpus and the
exploratory corpus without English.

\textbf{Table 2. M1 variance components for all languages and after
excluding English.}

{\def\LTcaptype{table} 
\begin{longtable}[]{@{}
  >{\raggedright\arraybackslash}p{(\linewidth - 8\tabcolsep) * \real{0.1579}}
  >{\raggedleft\arraybackslash}p{(\linewidth - 8\tabcolsep) * \real{0.2105}}
  >{\raggedleft\arraybackslash}p{(\linewidth - 8\tabcolsep) * \real{0.2105}}
  >{\raggedleft\arraybackslash}p{(\linewidth - 8\tabcolsep) * \real{0.2105}}
  >{\raggedleft\arraybackslash}p{(\linewidth - 8\tabcolsep) * \real{0.2105}}@{}}
\toprule\noalign{}
\begin{minipage}[b]{\linewidth}\raggedright
Component
\end{minipage} & \begin{minipage}[b]{\linewidth}\raggedleft
All: absolute variance
\end{minipage} & \begin{minipage}[b]{\linewidth}\raggedleft
All: share
\end{minipage} & \begin{minipage}[b]{\linewidth}\raggedleft
No English: absolute variance
\end{minipage} & \begin{minipage}[b]{\linewidth}\raggedleft
No English: share
\end{minipage} \\
\midrule\noalign{}
\endhead
\bottomrule\noalign{}
\endlastfoot
Brand & 0.0006071351 & 1.462916\% & 0.0003238816 & 2.234577\% \\
Language & 0.0109922979 & 26.486382\% & 0.0007341950 & 5.065478\% \\
Provider & 0.0006805145 & 1.639727\% & 0.0000863706 & 0.595902\% \\
Prompt & 0.0004411379 & 1.062939\% & 0.0001159274 & 0.799825\% \\
Residual & 0.0287806116 & 69.348035\% & 0.0132337168 & 91.304218\% \\
\textbf{Total fitted variance} & \textbf{0.0415016971} & \textbf{100\%}
& \textbf{0.0144940914} & \textbf{100\%} \\
\end{longtable}
}

In the full corpus, language accounts for 26.49\% of fitted variance and
brand accounts for 1.46\%. After English is excluded, the language share
is 5.07\% and the brand share is 2.23\%. The rise in the brand
percentage occurs while absolute brand variance falls from 0.0006071351
to 0.0003238816, a decrease of 46.65\%.

Total fitted variance falls from 0.0415016971 to 0.0144940914, a
decrease of 65.08\%. Absolute language variance falls by 93.32\%.
Provider, prompt, and residual variance also fall in absolute terms. The
new shares are ratios to a much smaller total. For this reason, the
increase in brand share does not indicate stronger absolute brand
separation in the non-English data.

The sensitivity result is large in descriptive terms. It shows that the
original 26.49\% language share is not stable to removal of English.
Subtracting 5.07 from 26.49 would not yield a valid additive English
component because the second model refits all components on a changed
sample.

As an exploratory specificity check, the same M1 specification was
refitted after omitting each of the other languages in turn. The fitted
language shares were 26.81\% without Czech, 27.97\% without Estonian,
27.27\% without Hungarian, 27.31\% without Lithuanian, 27.14\% without
Latvian, 27.62\% without Polish, and 27.29\% without Slovak. These
values remain close to or slightly above the 26.49\% full-corpus
estimate, while the English-exclusion value is 5.07\%. BFGS did not
converge for the Czech-exclusion fit; a Powell retry converged and
supplies the reported point estimate. This leave-one-language-out
comparison is descriptive and has no confidence intervals or
significance test.

\begin{figure}
\centering
\pandocbounded{\includegraphics[width=\linewidth,keepaspectratio,alt={Language sensitivity of the stored classifier outcome. Panel A reports the percentage of answers receiving a negative classifier score in each stored language. Panel B reports the fitted M1 language-component share for the full corpus and each leave-one-language-out refit. Values are classifier counts and model point estimates, without confidence intervals. They do not estimate causal language effects or human-coded brand sentiment.}]{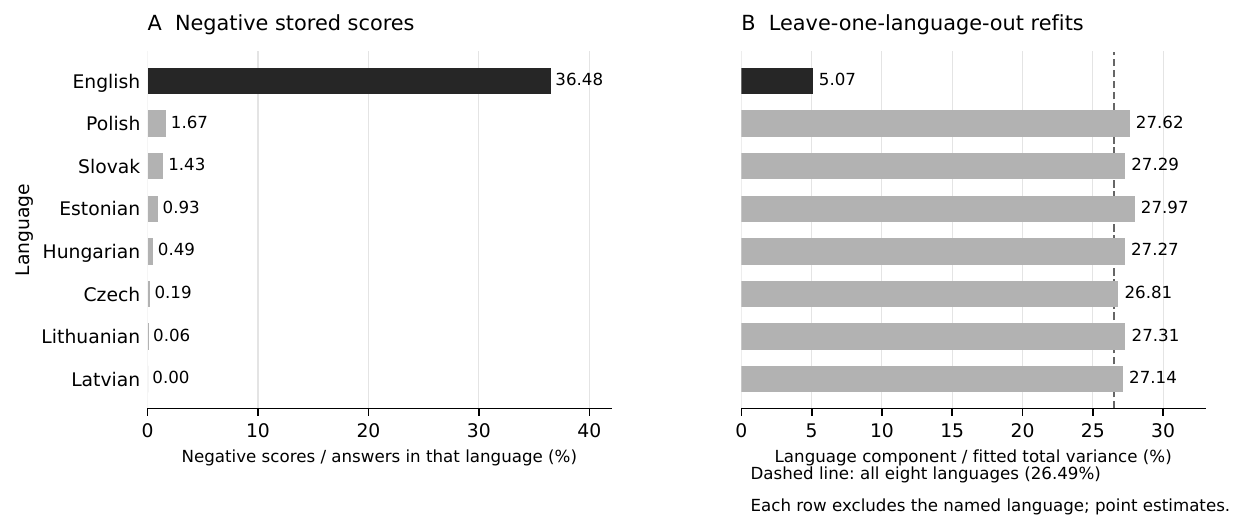}}
\caption{Language sensitivity of the stored classifier outcome. Panel A
reports the percentage of answers receiving a negative classifier score
in each stored language. Panel B reports the fitted M1
language-component share for the full corpus and each
leave-one-language-out refit. Values are classifier counts and model
point estimates, without confidence intervals. They do not estimate
causal language effects or human-coded brand sentiment.}
\end{figure}

\section{Discussion}\label{discussion}

\subsection{What the sensitivity analysis
establishes}\label{what-the-sensitivity-analysis-establishes}

The full-corpus M1 point estimates reproduce exactly, and the
English-exclusion point estimates reproduce across an independent
automated rerun. The central empirical conclusion is therefore narrow
and clear: the variance decomposition of this automated sentiment score
is highly sensitive to whether English is included.

The class counts identify where much of the score distribution differs.
Negative classifier scores occur in 36.48\% of English answers and
0.68\% of answers in the other seven languages. The same table also
shows positive-score differences, including 214 positive scores in
Estonian. M1 summarizes the entire numeric score, not the negative class
alone, so the negative-count comparison is a diagnostic description
rather than a decomposition of the language variance.

Several mechanisms remain compatible with the observed data. English
answers may differ in content, tone, structure, or length. The
classifier may behave differently across languages because of
fine-tuning coverage, calibration, or tokenization. The tweet domain may
transfer unequally to long-form company answers. The first 2,000
characters and 512-token truncation may expose different parts of
differently structured responses. The current analysis cannot separate
these mechanisms.

\subsection{Implications for automated
measurement}\label{implications-for-automated-measurement}

The result supports a measurement-first workflow. Before interpreting an
automated language difference as a difference in company reputation,
researchers should define the target construct and check how the
instrument represents it. For brand-focused sentiment, the unit of
annotation should identify which entity a statement concerns. A
response-level classifier with no target input can assign a score based
on language about competitors, an industry, uncertainty, or other
content in the answer.

A useful validation study would draw a blinded, stratified sample across
languages, prompt tasks, providers, and score classes. Qualified readers
would apply written rules that distinguish target-brand sentiment,
recommendation, criticism, and unclassifiable content. Multiple
independent coders would permit agreement assessment. The automated
score could then be compared with those labels by language and answer
type. No such new human coding was conducted for this correction.

The same caution applies to study design decisions. M1's residual
includes interactions and within-cell variation, so its 69.3\% or 91.3\%
residual share is not an estimate of pure generation noise. The three
repeated Category D prompts ask different questions. An analysis of
repeat reliability or query-budget allocation would need a model suited
to those tasks, a validated outcome, and uncertainty for the projected
designs. The corrected paper therefore does not retain the original
empirical M2 and M3 allocation and reliability tables.

\subsection{Revision from the earlier
manuscript}\label{revision-from-the-earlier-manuscript}

The earlier manuscript reported additional variance decompositions and a
decision-study allocation based on the Category D subset. Those results
are omitted from this correction, which retains the reproduced M1 point
estimates and exploratory language-exclusion sensitivity. The allocation
projected average error reduction over unequal query blocks,
extrapolated beyond the three observed Category D prompt levels, and
supplied no uncertainty intervals. These limitations prevent its use
here as evidence for a general rule about how many repeats, prompts,
providers, or languages to purchase.

This revision also removes claims about unrelated corporate and
single-brand studies, prospective bootstrap or preregistration work, and
any interpretation of the outcome as a recommendation or brand ranking.
The original archived manuscript remains a separate historical version.

\section{Limitations}\label{limitations}

First, the outcome has no multilingual human gold standard for the
target construct. The analysis concerns stored classifier scores. It
cannot determine whether an answer is objectively positive or negative
toward the named company.

Second, the score is zero-heavy: 91.85\% of observations are exactly
zero by construction when neutral wins. Empirical score variance differs
sharply by language. A Gaussian model with homogeneous residual variance
is a first-order summary and may fit the distribution poorly.

Third, eight language levels in the full model, seven in the sensitivity
model, and three provider levels provide limited information for
random-component estimation. The optimization emitted boundary and
numerical warnings. Point convergence does not supply calibrated
uncertainty.

Fourth, provider aliases do not identify immutable runtime model
versions. Reported collection settings include temperature 0.3, but
per-request receipts, finish reasons, and token use are unavailable. The
provider component should be read at the alias level recorded in the
dataset.

Fifth, the design is incomplete in a small number of repeated cells.
Category D contains 1,433 cells with five responses and two with four;
five planned cells have no retained response. M1 also pools prompt tasks
that differ in purpose and branding structure.

Sixth, the English-exclusion analysis is a sample-composition
sensitivity test. It does not estimate a causal effect of English and
cannot attribute the difference to the scorer alone.

Seventh, the brands were selected from a study of companies operating in
Central and Eastern European markets. Results need not generalize to
other brands, markets, time periods, provider systems, prompts, or
scoring instruments.

\section{Conclusion}\label{conclusion}

The original main-effects decomposition is reproducible as a point
estimate, but its largest systematic component depends strongly on
English inclusion. Query language accounts for 26.49\% of fitted
variance across all eight languages and 5.07\% after English is removed
in an exploratory refit. Total fitted variance falls by 65.08\%, and
absolute brand variance also falls even though its percentage share
rises. English contains 591 negative classifier scores among 1,620
answers; the other seven languages contain 77 among 11,313.

These results diagnose sensitivity in an automated measurement pipeline.
They do not establish that English answers express more negative brand
sentiment. The scorer's tweet-domain fine-tuning, language coverage,
truncation, and lack of a target-brand condition leave answer
differences and measurement differences unresolved. Substantive claims
about multilingual brand sentiment require a construct-specific,
multilingual validation sample.

\section{Declarations}\label{declarations}

\subsection{Funding}\label{funding}

This study was conducted at Rankfor.AI and received no external funding.

\subsection{Competing interests}\label{competing-interests}

The author is affiliated with Rankfor.AI Sp. z o.o., a commercial
company that provides AI visibility measurement services. This
relationship may constitute a commercial interest in methods for
measuring brand representation in LLM outputs.

\subsection{Ethics statement}\label{ethics-statement}

The study analyses outputs generated by provider systems in response to
research prompts. It contains no human participants or private personal
data. The correction added no human label coding.

\subsection{AI assistance disclosure}\label{ai-assistance-disclosure}

AI-assisted tools were used for code checking, automated statistical
reproduction, and drafting support. The author accepts responsibility
for the analysis, interpretation, and text. Automated review did not
constitute human validation of the classifier labels.

\subsection{Data and code
availability}\label{data-and-code-availability}

The aggregate parent dataset is available at Zenodo, DOI
\href{https://doi.org/10.5281/zenodo.20794390}{10.5281/zenodo.20794390}.
The response-level extract can be requested from the author, subject to
the source data terms and a suitable transfer route. Analysis code
accompanies this revision as an attachment. This statement does not
claim that the response-level extract or revision code has been
deposited publicly.

\section{References}\label{references}

\begin{enumerate}
\def\labelenumi{\arabic{enumi}.}
\tightlist
\item
  Barbieri, F., Espinosa Anke, L., and Camacho-Collados, J. (2022).
  XLM-T: Multilingual language models in Twitter for sentiment analysis
  and beyond. \emph{Proceedings of LREC 2022}, 258-266.
  \href{https://aclanthology.org/2022.lrec-1.27/}{ACL Anthology
  2022.lrec-1.27}.
\item
  Conneau, A., Khandelwal, K., Goyal, N., Chaudhary, V., Wenzek, G.,
  Guzmán, F., Grave, E., Ott, M., Zettlemoyer, L., and Stoyanov, V.
  (2020). Unsupervised cross-lingual representation learning at scale.
  \emph{Proceedings of ACL 2020}, 8440-8451.
  \href{https://doi.org/10.18653/v1/2020.acl-main.747}{doi:10.18653/v1/2020.acl-main.747}.
\item
  Cardiff NLP. (n.d.).
  \nolinkurl{cardiffnlp/twitter-xlm-roberta-base-sentiment-multilingual}
  model card. Hugging Face. Accessed 1 November 2026.
  \url{https://huggingface.co/cardiffnlp/twitter-xlm-roberta-base-sentiment-multilingual}.
\item
  Cardiff NLP. (n.d.). Tweet sentiment multilingual dataset card.
  Hugging Face. Accessed 1 November 2026.
  \url{https://huggingface.co/datasets/cardiffnlp/tweet_sentiment_multilingual}.
\item
  Atıl, B., et al.~(2025). Non-determinism of ``deterministic'' LLM
  system settings in hosted environments. \emph{Proceedings of the 5th
  Workshop on Evaluation and Comparison of NLP Systems}, 135-148.
  \href{https://doi.org/10.18653/v1/2025.eval4nlp-1.12}{doi:10.18653/v1/2025.eval4nlp-1.12}.
\item
  Ouyang, S., Zhang, J. M., Harman, M., and Wang, M. (2025). An
  empirical study of the non-determinism of ChatGPT in code generation.
  \emph{ACM Transactions on Software Engineering and Methodology},
  34(2). \href{https://doi.org/10.1145/3697010}{doi:10.1145/3697010}.
\item
  Cronbach, L. J., Gleser, G. C., Nanda, H., and Rajaratnam, N. (1972).
  \emph{The Dependability of Behavioral Measurements: Theory of
  Generalizability for Scores and Profiles}. Wiley.
\item
  Shavelson, R. J., and Webb, N. M. (1991). \emph{Generalizability
  Theory: A Primer}. Sage.
\item
  Searle, S. R., Casella, G., and McCulloch, C. E. (1992).
  \emph{Variance Components}. Wiley.
  \href{https://doi.org/10.1002/9780470316856}{doi:10.1002/9780470316856}.
\item
  Bates, D., Mächler, M., Bolker, B., and Walker, S. (2015). Fitting
  linear mixed-effects models using lme4. \emph{Journal of Statistical
  Software}, 67(1), 1-48.
  \href{https://doi.org/10.18637/jss.v067.i01}{doi:10.18637/jss.v067.i01}.
\item
  Seabold, S., and Perktold, J. (2010). statsmodels: Econometric and
  statistical modeling with Python. \emph{Proceedings of the 9th Python
  in Science Conference}, 92-96.
  \href{https://doi.org/10.25080/Majora-92bf1922-011}{doi:10.25080/Majora-92bf1922-011}.
\item
  Żatuchin, D. (2026). The synthesised consideration set: brand
  inclusion and concentration in large language model recommendations
  across five industries and three systems. \emph{Journal of Marketing
  Analytics}, published online 31 October 2026.
  \href{https://doi.org/10.1057/s41270-026-00562-3}{doi:10.1057/s41270-026-00562-3}.
\item
  Żatuchin, D. (2026). The Dice Roll Method: A Standardized Protocol for
  Repeated-Query Auditing of Large Language Model Brand Recommendations.
  arXiv:2609.04047v2, revised 14 October 2026.
  \url{https://arxiv.org/abs/2609.04047v2}.
\item
  Żatuchin, D. (2026). \emph{Category Ownership Map: Corrected Analyses
  and September Replication Outputs (2026)}, version 3.0, 24 September
  2026. Zenodo.
  \href{https://doi.org/10.5281/zenodo.22927143}{doi:10.5281/zenodo.22927143}.
\item
  Żatuchin, D. (2026). \emph{Cross-Language AI Brand Reputation: A
  66-Brand, 12-Language Dataset}, version 1, 22 June 2026. Zenodo.
  \href{https://doi.org/10.5281/zenodo.20794390}{doi:10.5281/zenodo.20794390}.
\end{enumerate}
\end{document}